\documentclass[amsmath, amssymb, aps, longbibliography, prl, superscriptaddress,
preprint,
nobibnotes, 
]{revtex4-2}

\makeatletter
\def\NAT@spacechar{}
\makeatother

\usepackage[T1]{fontenc}
\usepackage{amsmath}
\usepackage{mathtools}
\usepackage[nolimits]{cmupint}
\usepackage[separate-uncertainty=true]{siunitx}
\usepackage{physics}
\AtBeginDocument{\RenewCommandCopy\qty\SI}
\ExplSyntaxOn
\msg_redirect_name:nnn { siunitx } { physics-pkg } { none }
\ExplSyntaxOff
\DeclareSIUnit{\photons}{\ensuremath{\mathrm{photons}}}
\newcommand{\qqq}[1]{\ensuremath{\quad\quad\text{#1}\quad\quad}}
\usepackage[colorlinks=true]{hyperref}
\AtBeginDocument{\definecolor{linkcolor}{rgb}{0,0.4470,0.7410}\hypersetup{allcolors=linkcolor}}
\usepackage{graphicx}
\usepackage{tabularx}
\usepackage{booktabs}
\usepackage{enumitem}
\usepackage{diagbox}
\usepackage{mathrsfs}

\let\oldfrac\frac
\renewcommand{\frac}[2]{%
  \mathchoice
    {\oldfrac{#1}{#2}}
    {#1/#2}
    {#1/#2}
    {#1/#2}
}

\let\oldautoref\autoref{}

\renewcommand{\autoref}[1]{%
    \begingroup%
    \def\figureautorefname{Figure}%
    \def\tableautorefname{Table}%
    \def\partautorefname{Part}%
    \def\appendixautorefname{Appendix}%
    \def\equationautorefname{Equation}%
    \def\itemautorefname{Item}%
    \def\chapterautorefname{Chapter}%
    \def\sectionautorefname{Section}%
    \def\subsectionautorefname{Subsection}%
    \def\subsubsectionautorefname{Subsubsection}%
    \def\paragraphautorefname{Paragraph}%
    \def\Hfootnoteautorefname{Footnote}%
    \def\AMSautorefname{Equation}%
    \def\theoremautorefname{Theorem}%
    \oldautoref{#1}%
    \endgroup%
}

\newlength{\plotwidth}

\begin{document}

\definecolor{linkcolor}{rgb}{0,0.4470,0.7410}%
\hypersetup{allcolors=linkcolor}

\title{Nonlinear reversal of photo-excitation on the attosecond time scale improves ultrafast x-ray diffraction images}

\newcommand{\TUB}{\affiliation{Institute of Optics and Atomic Physics, Technische Universit\"at Berlin, 10623 Berlin, Germany}}
\newcommand{\UHH}{\affiliation{CFEL, Universit\"at Hamburg, 22761 Hamburg, Germany}}
\newcommand{\ETH}{\affiliation{ETH Zurich,Laboratory of Solid State Physics, 8093 Zurich, Switzerland}}
\newcommand{\EuXFEL}{\affiliation{European XFEL, 22869 Schenefeld, Germany}}
\newcommand{\DESY}{\affiliation{Deutsches Elektronen-Synchrotron DESY, 22607 Hamburg, Germany}}
\newcommand{\LCLS}{\affiliation{Linac Coherent Light Source, SLAC National Accelerator Laboratory, Menlo Park, CA 94025, USA}}
\newcommand{\SLAC}{\affiliation{SLAC National Accelerator Laboratory, Menlo Park, CA 94025, USA}}
\newcommand{\PULSE}{\affiliation{Stanford PULSE Institute, SLAC National Accelerator Laboratory, Menlo Park, CA 94025, USA}}
\newcommand{\Stanford}{\affiliation{Physics Department, Stanford University, Stanford, CA 94025, USA}}
\newcommand{\StanfordApplied}{\affiliation{Department of Applied Physics, Stanford University, Stanford, CA, USA}}
\newcommand{\Argonne}{\affiliation{Argonne National Laboratory, Argonne, IL 60439, USA}}
\newcommand{\PSI}{\affiliation{PSI, Paul-Scherrer-Institute, 5232 Villigen, Switzerland}}
\newcommand{\TUD}{\affiliation{Technical University Darmstadt, Institute of Nuclear Physics, Schlossgartenstr. 9, 64289 Darmstadt, Germany}}
\newcommand{\EPFL}{\affiliation{
Laboratory for Ultrafast X-ray Sciences, Institute of Chemical Sciences and Engineering (LUXS), École Polytechnique Fédérale de Lausanne (EPFL), Lausanne, Switzerland}}

\author{Anatoli~Ulmer}
    \email[Corresponding author: ]{anatoli.ulmer@desy.de}\UHH{}
\author{Phay~J.~Ho}\Argonne{}
\author{Bruno~Langbehn}\TUB{}
\author{Stephan~Kuschel}\UHH{}\TUD
\author{Linos~Hecht}\ETH{}
\author{Razib~Obaid}\SLAC{}
\author{Simon~Dold}\EuXFEL{}
\author{Taran~Driver}\SLAC{}\PULSE{}
\author{Joseph~Duris}\SLAC{}
\author{Ming-Fu~Lin}\SLAC{}
\author{David~Cesar}\SLAC{}
\author{Paris~Franz}\SLAC{}
\author{Zhaoheng~Guo}\SLAC{}\StanfordApplied
\author{Philip~A.~Hart}\SLAC{}
\author{Andrei~Kamalov}\SLAC{}
\author{Kirk~A.~Larsen}\SLAC{}
\author{Xiang~Li}\SLAC{}
\author{Michael~Meyer}\EuXFEL{}
\author{Kazutaka~Nakahara}\SLAC{}
\author{Robert~G.~Radloff}\UHH{}
\author{River Robles}\SLAC{}
\author{Lara~R\"onnebeck}\UHH{}
\author{Nick~Sudar}\SLAC{}
\author{Adam~M.~Summers}\SLAC{}
\author{Linda~Young}\Argonne{}
\author{Peter~Walter}\SLAC{}
\author{James~Cryan}\SLAC{}\PULSE{}
\author{Christoph~Bostedt}\PSI\EPFL
\author{Daniela~Rupp}\ETH{}
\author{Agostino~Marinelli}\SLAC{}
\author{Tais~Gorkhover}
    \email[Corresponding author: ]{tais.gorkhover@cfel.de}\UHH{}

\date{\today}

\begin{abstract}
The advent of isolated and intense sub-femtosecond X-ray pulses enables tracking of quantum-mechanical motion of electrons in molecules and solids. The combination of X-ray spectroscopy and diffraction imaging is a powerful approach to visualize non-equilibrium dynamics in systems beyond few atoms. However, extreme x-ray intensities introduce significant electronic damage, limiting material contrast and spatial resolution. Here we show that newly available intense sub-femtosecond (sub-fs) x-ray FEL pulses can outrun most ionization cascades and partially reverse x-ray damage through stimulated x-ray emission in the vicinity of a resonance. In our experiment, we compared thousands of coherent x-ray diffraction patterns and simultaneously recorded ion spectra from individual Ne nanoparticles injected into the FEL focus. Our experimental results and theoretical modeling reveal that x-ray diffraction increases and the average charge state decreases in particles exposed to sub-fs pulses compared to those illuminated with 15-femtosecond pulses. Sub-fs exposures outrun most Auger decays and impact ionization processes, and enhance nonlinear effects such as stimulated emission, which cycle bound electrons between different states. These findings demonstrate that intense sub-fs x-ray FEL pulses are transformative for advancing high-resolution imaging and spectroscopy in chemical and material sciences, and open the possibilities of coherent control of the interaction between x-rays and complex specimen beyond few atoms.

\end{abstract}

\maketitle
The information content of virtually all x-ray images of objects, ranging from tens of centimeters down to the atomic scale, is limited by photo-ionization damage induced throughout the exposure. Usually, the x-ray dose has to be increased in order to obtain a brighter image with higher contrast and better spatial resolution. The drawback of a higher x-ray fluence is the enhanced rate of ionization events, which is detrimental to the integrity of the sample. The maximum allowable dose before significant modification of the specimen occurs is a hard limit to the signal-to-noise ratio of the image. This applies even to advanced schemes such as diffraction-before-destruction imaging \cite{Neutze2000,Seibert2011,Barty2012,Yumoto2022}. Here, intense x-ray FEL flashes as short as a few femtoseconds outrun the ionic structure damage and provide unique insights into non-equilibrium dynamics in molecules and other nanoscale specimen with unprecedented temporal and spatial precisions \cite{Barke2015,Loh2012, Schriber2022, Bostedt2012,Gorkhover2012,Rupp2020,Gomez2014,Gorkhover2016, Ihm2019direct, Ferguson2016transient, Peltz2022few}. A significant limitation of this approach remains the electronic bleaching of the sample \cite{Neutze2000, Ho2017, Son2011impact, Schropp2010, Aquila2015}. This applies to both, imaging and spectroscopic experiments based on intense x-ray flashes. After photoionization, a damage cascade through Auger decays and other relaxation processes occurs already during the first few femtoseconds. In the extreme case, x-rays can produce ions without bound electrons turning them transparent within few femtoseconds before the sample degrades on the ionic structure level. This is regarded as the ultimate limit to the spatial resolution in x-ray diffraction imaging and is also a limit for signal-to-noise optimization in spectroscopic studies, especially for site-specific nonlinear schemes in complex specimen \cite{Young2010, Aquila2015, Barty2012, Jurek2008effect, Lorenz2012impact, Ferguson2016transient, Ho2017, Kroll2018Stimulated,Ho2020, Rohringer2012XrayLaser, Eichmann2020Recoil}. 

The advent of high peak power TW 300--500 attosecond short x-ray pulses provides an interesting route to beat most electronic damage mechanisms \cite{Duris2020,Li2022attosecond,Franz2024terawatt,Driver2024}. Short, sub-fs, intense x-ray bursts are faster than Auger decay times of light materials, which are around 2--3\,fs. Also, at high x-ray peak powers nonlinear effects gain in probability and significance \cite{Rohringer2008resonant,Rohringer2012XrayLaser,Eichmann2020Recoil}. In our study, we find that the quality of x-ray diffraction images from Ne nanoparticles can be increased using intense sub-fs x-ray pulses tuned to the Ne K-edge. We compared sub-fs short exposures with snapshots recorded with 15\,fs FEL pulses. Both exposure times are short enough to outrun the ionic damage. First, we observe that transient resonances can increase the scattering cross sections beyond literature values. Second, the overall energy absorbed from the sub-fs x-ray pulse after the imaging process is lower compared to few-femtosecond exposures. We also find that partial x-ray damage reversal through stimulated emission significantly increases the brightness of the images. Third, our simulation indicates that the portion of diffraction signal from bound electrons is significantly higher in snapshots recorded with sub-fs pulses compared to 15\,fs exposures. Longer pulses reflect the formation of a hot and dense nanoplasma where the ions appear as frozen, but many electrons contributing to x-ray scattering are delocalized. Sub-fs x-ray pulses outrun most nanoplasma formation steps and thus, the resulting diffraction patterns rather mirror the pristine sample.

\section*{Measurement}

In our study, we explore the scattering cross sections and x-ray absorption yields of individual Ne nanoparticles as a function of FEL pulse parameters such as FEL photon energy $\hbar\omega$, FEL fluence and pulse duration $\tau$ \cite{Duris2020, Driver2024}. For $\tau\approx\SIrange{0.3}{0.5}{fs}$, we have collected thousands of x-ray diffraction patterns and simultaneously recorded ion spectra at x-ray photon energies tuned to the vicinity of the Ne K-edge at 870\,eV. We have also recorded a reference data set with $\tau\approx15\,$fs, which outruns ionic structure damage on the nanometer scale, but reflects the impact of electronic bleaching processes such as Auger decay and collisional ionization during nanoplasma formation \cite{Wabnitz2002,Gorkhover2012,Bostedt2012,Gorkhover2016}. The x-ray fluence was similar for both pulse durations (see Supplements).

A detailed schematic of our experiment is exhibited in \autoref{fig:exp}. Individual near-spherical Ne nanoparticles with diameters of \SIrange{60}{100}{nm} intersect the path of focused and intense single x-ray FEL pulses inside the LAMP endstation at the Linac Coherent Light Source (LCLS) \cite{Emma2010,Osipov2018lamp, Walter2022TMO}. Single-particle, single-exposure x-ray diffraction snapshots are recorded at 60\,Hz using the ePix dectector \cite{ePix}, which is located \SI{0.395}{m} downstream from the interaction region.  The particle size is encoded directly into the Airy pattern-like diffraction patterns and can be recovered with $0.3$\,nm accuracy \cite{Guinier1955, Gorkhover2016}. Each diffraction image is correlated to a simultaneously recorded ion time-of-flight spectrum, which carries an imprint of the total photoenergy absorbed by the nanoparticle \cite{Gorkhover2012}. Shorter flight times indicate higher ion energies. Both detectors and the Ne nanoparticle source are synchronized to the FEL pulse arrival times. (see Supplements, Experimental setup for details). 

We scan the photon energy $\hbar\omega$ in the vicinity of the Ne K-shell absorption edge between $815\,$eV and $1000\,$eV. Each scanning step contains thousands of diffraction patterns from individual Ne nanoparticles with fluctuating brightness due to random positions inside the \SI{1.2}{\micro\meter} FEL focus with a near-Gaussian x-ray fluence distribution \cite{Gorkhover2012} (see Supplements, Experimental setup). 

\begin{figure}[ht]
    \centering
    \includegraphics[width=1\textwidth]{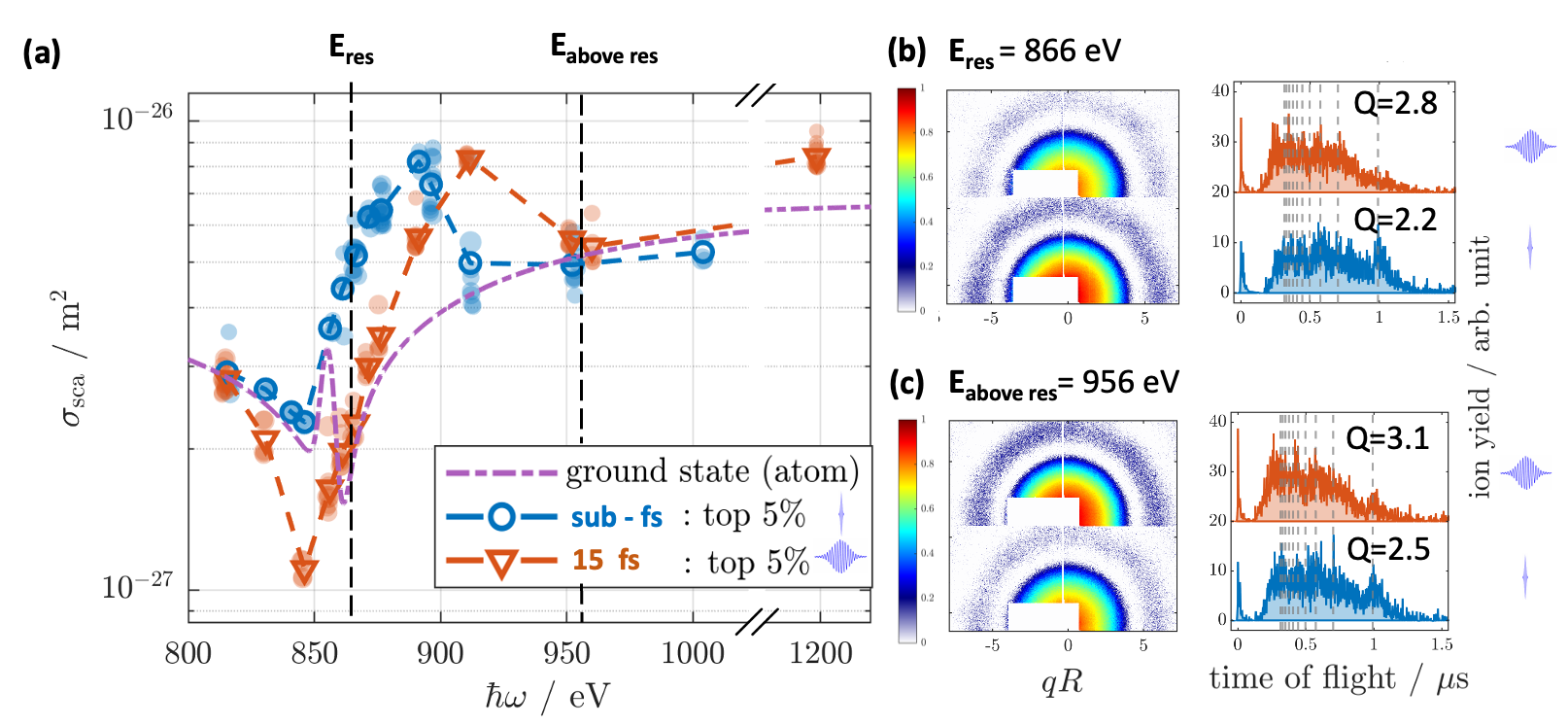}

    \caption{Diffraction and absorption trends in the vicinity of the Ne K-edge for sub-fs  (blue) and 15\,fs (red) exposures. Panel (a) exhibits the measured scattering cross sections $\sigma_\mathrm{sca}$ of the brightest 5 percent of all images per photon energy $\hbar\omega$ (x-axis). The measured scattering cross cross sections are amplified beyond the neutral state (purple dashed line) for certain energies for both pulse durations. However, the absorption increases less for sub-fs pulses. In panel (b) and (c), eight of the brightest diffraction patterns (b,left side) and the simultaneously recorded ion time-of-flight spectra (b, right side) are shown. The absorbed energy can be measured through ion counts (y-axis) per ion time-of-flight (x-axis). The shorter the average time-of-flight, the more energy was absorbed. We converted the average time-of-flight into the corresponding average charge state Q (by neglecting the kinetic energy of ions). Ne charge states are the dotted vertical lines inside the ion spectra.  Near the Ne K-edge resonance at 866\,eV, the diffraction patterns for sub-fs pulses are brighter, but the corresponding average charge state is smaller compared to the 15\,fs exposure (panel (b)). Above the resonance, this ratio is reversed. The brightest average image is connected to a higher average charge state. The diffraction patterns are plotted over the product between the scattering vector \textit{q} and the nanoparticle radius \textit{R}.}
        \label{fig:single}
\end{figure}

The measured variation of the absorbed energy and the diffraction cross sections across the photon energy scan is exhibited in \autoref{fig:single}. From each individual diffraction pattern, we extracted the average scattering cross section per Ne atom inside the nanoparticle (see Supplements, Data processing). The top 5\% of the measured scattering cross sections for Ne atoms $\sigma_\mathrm{sca}$ are plotted against the incoming photon energy $\hbar\omega$. Each dot corresponds to a single image, blue dots indicate data recorded with sub-fs pulses and red with 15\,fs pulses. The observed cross sections align overall with literature values for neutral Ne atoms (purple dashed line) except near the Ne K-shell resonance at 870\,eV. Here, $\sigma_\mathrm{sca}$ is amplified by a factor of 2 to 3 compared to the neutral Ne curve for both pulse durations. The highest increase is close to 900\,eV. Such behavior was already attributed to transient resonances in a previous study in Xe nanoparticles \cite{Kuschel2025}.  The intense FEL pulses create new ionic states, which can become resonant to the incoming x-ray photon energy. In the Xe nanoparticle study, the enhanced x-ray diffraction came at the cost of increased absorption efficiency for all pulse durations.

Our experimental data paints a more complex picture of the absorption vs diffraction trends as displayed in panels (b) and (c) of \autoref{fig:single}. In panel (b) and (c), an average of eight of the brightest snapshots (most scattered photons per atom) are depicted per $\tau$ at two selected photon energies. The corresponding x-ray diffraction patterns (left) and their respective average ion spectra (right) recorded at photon energy $E_\mathrm{res} = 866\,$eV with  $\tau \approx15$\,fs  (b, top) and $\tau \approx0.3$\,fs pulses (b, bottom) are displayed. The diffraction image from the sub-fs pulse is brighter than that from the longer exposure. At the same time the nanoparticles have absorbed less energy from the shortest pulse as witnessed by a lower average charge state $Q$ (b, top right). This unusual behavior seems to disappear further above the K-edge at $E_\mathrm{above\:res} = 956\,$eV (see the panels in (c)). Here, the shorter exposure is also accompanied by a lower $Q$ compared to the 15\,fs exposure. In parallel, the x-ray diffraction recorded by the sub-fs pulse is dimmer, which suggests that increased scattering signal is linked to enhanced absorption similarly to previous observations \cite{Kuschel2025}. Please note that $Q$ does not reflect the exact average charge state during the exposure, but is rather a relative measure of the total absorbed energy per atom \cite{Gorkhover2012, Arbeiter2011rare}.

The selected images displayed in \autoref{fig:single} represent a trend, which emerges from the full data set  shown in \autoref{fig:statistics}. Here, the correlation between the photon numbers scattered per Ne atom inside the nanoparticle (x-axis) is plotted versus the average charge states per ion $Q$ extracted from the coincident ion spectra (y-axis). Each dot is a single  FEL exposure with notable diffraction signal from a single nanoparticle.  We calculated the number of photons scattered per atom from the radial profile fits of the scattering patterns and the Ne solid state density (see Supplements). The simultaneously recorded ion time-of-flight spectra mirror the fraction of the FEL pulse energy absorbed by the nanoparticle. More absorbed x-ray photons result in a higher average charge state $Q$ detected nanoseconds after the x-ray pulse. The larger dots are the eight brightest shots shown in \autoref{fig:single}. 

Panel (a) of \autoref{fig:statistics} depicts the data recorded at $E_\mathrm{res} = 866\,$eV and panel (b) at $E_\mathrm{above\:res} = 956\,$eV, both recorded with sub-fs and 15\,fs pulse durations. The FEL fluence distribution inside the FEL focus causes a wide spread of absorbed and diffracted x-ray energies. The 60--100\,nm small particles are randomly injected into the much larger focal spot of the FEL with a full width at half maximum of around \SI{1}{\micro\meter}. Thus, nanoparticles experience a broad variation of FEL fluences based on their position inside the focal volume. 

The overall trend is similar for both pulse durations, the number of scattered photons per atom (x-axis) increases with the average charge state (y-axis). However, the gradients of this trend diverge for sub-fs and 15\,fs pulses at $E_\mathrm{res}$ and converge again at $E_\mathrm{above\:res}$. 
In panel (a) recorded at $E_\mathrm{res}$, the average charge state increases faster for 15\,fs exposures without a significant enhancement of the scattering signal. In contrast, sub-fs exposures recorded near the FEL focus center witness more scattered photons per atom while $Q$ remains relatively moderate even in the brightest shots. In panel (b), which exhibits data recorded at $E_\mathrm{above\:res}$, the rises in the average charge state and the scattering are very similar for both x-ray pulse durations.

Near Ne K-edge we observe in \autoref{fig:single} and \autoref{fig:statistics}, that sub-fs FEL pulses are less efficiently absorbed but simultaneously more efficiently diffracted  compared to few-femtosecond FEL pulses with a similar photon number. This diverging behavior is surprising in the light of previous experimental studies. Amplified absorption through transient resonances has been demonstrated in the XUV/soft x-ray regime in the past \cite{Kanter2011,Rudek2012,Bostedt2012,Rupp2020,Rorig2023multiple,Kuschel2025}. In a previous study, transient resonances in oxygen atoms in sucrose nanospheres caused dimmer images because the experiment was carried out with $170-200$\,fs long pulses \cite{Ho2020}. The corresponding simulations suggested that transient resonances accelerate the sample´s electronic and ionic damage, and thus lead to a degradation in images. A follow up study on Xe nanoparticles has demonstrated that diffraction can be amplified through transient resonances, but at a cost of increased absorption rates similarly for 15\,fs and sub-fs FEL pulse durations \cite{Kuschel2025}. The Xe nanoparticle study was recorded with significantly lower FEL fluences in sub-fs FEL pulses, thus the influence of nonlinear effects was overall less prominent.

\section*{Modeling}
\begin{figure}[ht]
    \centering
    \includegraphics[width=1\textwidth]{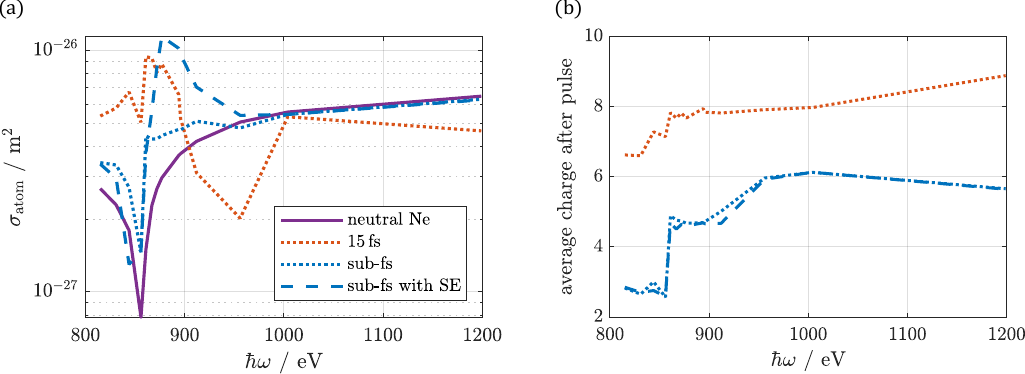}

    \caption{Simulation results. We calculated the expected scattering cross section per atom inside the Ne nanoparticle exposed to FEL pulse conditions similar to the experimental focus center. The electronic damage after the FEL pulse is mirrored by the average charge state inside the nanoparticle right after the FEL pulse, as exhibited in panel (b). The calcuations are plotted according to the incoming FEL photon energy (x-axis). Dashed lines are simulations with stimulated emission, dotted lines without. Blue lines represent sub-fs exposures, red lines show 15\,fs exposures. For interpretation see text and Supplements, Theory }
       \label{fig:theory}
\end{figure}

We  used our Monte Carlo-based calculation of nanoplasma effects inside Ne nanoparticles to model our experimental observations  \cite{Ho2016, Ho2017, Ho2014, Ho2020, Kuschel2025}. The simulation results are displayed in \autoref{fig:theory}. Our calculations indicate that sub-femtosecond pulses with energies tuned to transient resonances, can send excited electrons back into the K-shell through stimulated emission. Thus, parts of the absorbed energy from the x-ray pulse are re-emitted back into the light field. This de-excitation mechanism can become relevant if the pulses are shorter than Auger-Meitner decays and nanoplasma formation, which irreversibly dissipate the energy absorbed through photo-ionization further into the sample. Stimulated emission has two effects during the FEL exposure. First, some x-ray damage inside the nanoparticle will be reversed. Second, the bound electrons remain longer resonant to the x-rays and scatter more efficiently throughout the exposure.

To test the hypothesis, we compare the elastic scattering cross sections and the average charge states of Ne nanoparticles, which are exposed to FEL pulses close to the experimental conditions. In \autoref{fig:theory} panel (a), the calculated scattering cross sections are plotted versus the incoming photon energies. Without stimulated emission, the scattering cross sections produced by 300 attosecond FEL pulses (dotted blue line) are slightly above the neutral Ne case (solid purple line). If Ne nanoparticles are illuminated by 15\,fs pulses, the scattering efficiency is elevated compared to the sub-fs exposure in the vicinity of the absorption edge and decreases for higher energies. This observation is reasonable as there is a high density of transient resonances from Ne$^{ 1+}$ to Ne$^{ 5+}$ states, which can enhance the scattering efficiency before ionic desintegration of the sample (see Supplements, Theory). Far above the edge, there are much fewer resonances and the amplification is suppressed. However, in the experiment we observe that sub-fs pulses exposures are scattered more efficiently in the vicinity of the of the K-edge than suggested by our first model.  If stimulated emission is included into the simulation (blue dashed line), the scattering cross sections rapidly exceed amplifications achieved through longer pulses. This happens because of a partial recycling of transient resonances states as electrons in resonant excited states are cycled between the excited and original ground states. This prolonged life time of resonant ground states amplifies the diffraction over multiple photon scattering events. 

The resonant states recycling and x-ray scattering amplification occur while the ionization state throughout the exposure remains comparatively low. The average charge state after the FEL irradiation ranges between 2+ and 4+ as depicted in \autoref{fig:theory} panel (b) by the blue dotted line. This corresponds to up to three photons absorbed per atom. In comparison, 15\,fs FEL exposures produce average charge states higher than Ne$^{ 6+}$ to Ne$^{ 8+}$ (see the red dotted line). The charge states are increased through Auger relaxation processes and nanoplasma formation effects. In fact, this means that a significant portion of electrons, which contribute to the scattering cross section, are delocalized from their parent ions. These electrons are confined in the space charge Coulomb potential, which rapidly builds up during the exposures. Their locations are not directly related to the positions of individual atoms. 

In contrast, enhanced resonant scattering of sub-fs pulses stems mostly from electrons which are bound to parent ions. These coherent diffraction patterns mirror the pristine ionic structure because ion movement can be regarded as frozen during sub-fs short exposures. According to our simulation, stimulated emission excited by sub-fs pulses beats electronic bleaching and further increases the scattering cross section even if the x-ray fluence is increased by an order of magnitude (See Supplements, \autoref{fig:Ne_scaling}). 


Overall, our study demonstrates that intense sub-fs FEL pulses can increase elastic x-ray diffraction efficiency in combination with decreased sample damage. Our simulation explains this behavior through two effects. First, sub-fs pulses outrun the Auger decay processes and most of nanoplasma formation effects. Thus, x-ray diffraction occurs mainly on bound electrons. Second, stimulated emission works in favor of diffraction enhancement through transient resonances as excited electrons can return to resonant states multiple times during the FEL exposure and thus, the scattering amplification becomes more likely. Interestingly, the overall charge state after the pulse is not greatly affected by stimulated emission as shown in \autoref{fig:theory} panel (b), dashed (stimulated emission included) vs dotted blue lines (no stimulated emission). On average only a small portion of all ions emit stimulated emission mostly at the peak of the pulse. This little fraction of ions suffices to increase the scattering cross sections significantly, because the scattering cross section of a single transient ion can be several orders of magnitude higher compared to the neutral Ne (see Supplements, Theory). The highest transient resonance for Ne$^{9+}$ is at 1600\,eV and would in theory be able to provide sub-nm resolution images. These findings shed light on the potential of ultrafast x-ray imaging with sub-nanometer resolution for light elements, which has been unlocked through the development of intense sub-fs x-ray FEL pulses. For heavier elements, sub-fs pulses in the hard x-ray regime can deliver similar effects at even shorter wavelengths. 

\section*{Outlook}
In summary, our study is a demonstration of new sample damage-control capabilities using intense sub-fs FEL pulses. Such short pulses in combination with extreme x-ray peak powers outrun and even partially reverse electronic damage. The study questions the current picture of the electron bleaching limit in FEL experiments. In theory damage reversal can occur through stimulated emission, but might be also coherently controlled through Rabi oscillations as sub-fs pulses are fully temporal coherent. In principle, Rabi oscillations in Ne$^+$ in the soft x-ray regime have been predicted in the past \cite{Rohringer2008resonant,Kanter2011} and demonstrated in the XUV regime \cite{Nandi2022Observation,Richter2024Rabi}. The results of our study can be applied to other light materials such as oxygen or nitrogen, which are relevant for organic chemistry. In combination with hard x-rays, sub-fs pulse durations could increase material contrast and spatial resolution in virtually all diffraction-before-destruction experiments. With the prospect to coherently control and limit x-ray damage during the imaging process, our study is a significant step towards direct visualization of chemical processes and ultrafast phase transitions with unprecedented precision.

\clearpage

\begin{figure}[ht]
    \centering
    \includegraphics[width=1\textwidth]{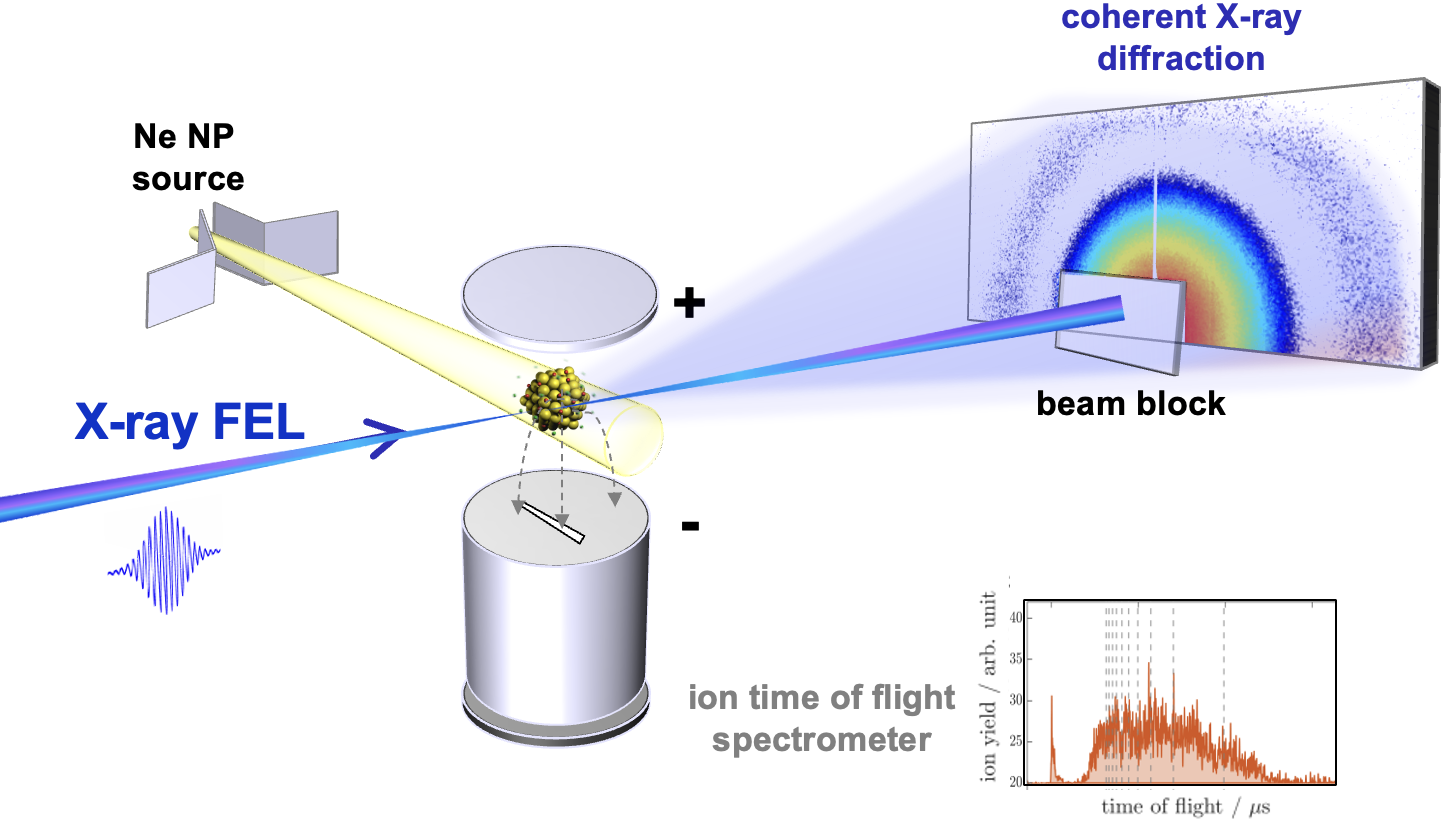}
  
    \caption{Experimental setup scheme: Individual Ne nanoclusters from a pulsed and cooled supersonic expansion Ne gas source are injected into focused the path of x-ray FEL flashes with different pulse lengths. Coherently diffracted photons were recorded by the two-dimensional x-ray detector ePIX, while a beam block protected the detector from the primary beam. An ion time-of-flight detector positioned at the FEL focus recorded cluster fragments from the sample expansion, which occurs nanoseconds after the FEL exposure. Please see Supplements, Experimental Details for more information.}
    
      \label{fig:exp}
\end{figure}

\begin{figure}[ht]
    \centering
    \includegraphics[width=1\textwidth]{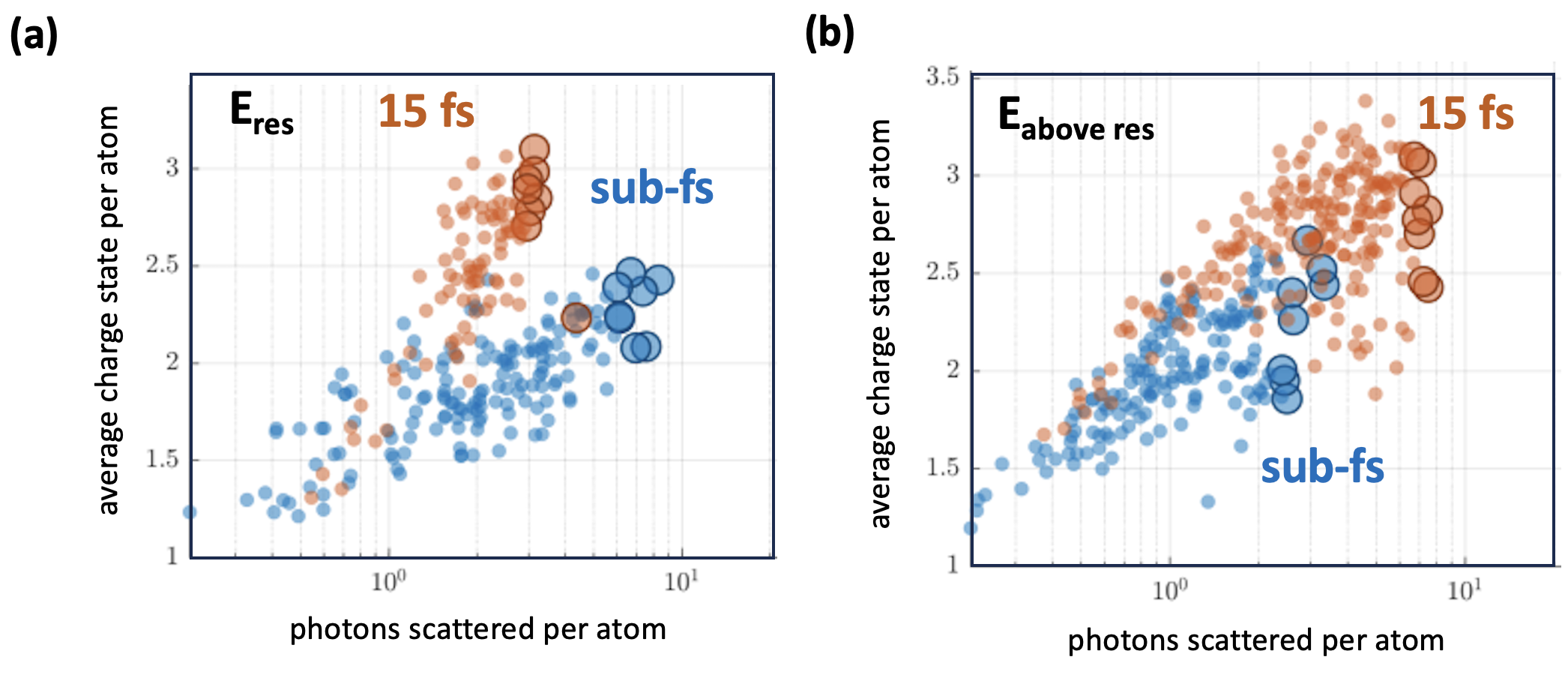}
       
    \caption{A statistical analysis for all shots recorded at $E_\mathrm{res}$ (panel(a)) and $E_\mathrm{above\:res}$ (panel(b)) is shown. Each dot represents a single measurement of a diffraction pattern and the corresponding ion spectrum. The larger
    dots represent the top 5 percent of shots depicted in \autoref{fig:single} for both aforementioned photon energies. In both panels, the photons scattered per Ne atom inside the  individual nanoparticle (x-axis) are plotted versus the average charge state per atom $Q$ (y-axis). At resonance depicated by the panel (a), the average charge state increases fast while the diffraction remains moderate for 15\,fs exposures. In contrast, the images recorded with sub-fs exposure demonstrate a higher number of photons scattered per atom while the average charge state increases much slower. The slopes of correlations converge above the reschosonance in panel (b). Please note, that the normalized number of photons per atoms refers to forward scattering from isolated atoms. The normalization includes the impact of the measured size of the particles as described in Supplements, Data Processing.}
     \label{fig:statistics}
\end{figure}

\clearpage

\section{Contributions}

T.G. conceived the idea for the experiment based on discussions with D.R., C.B. and A.M. A.U., B.L., D.R., designed and optimized the Ne droplet source. J.D., D.C. and A.G. implemented TW sub-fs pulses. All authors contributed to the experimental campaign. A.U. carried out data analysis with input from T.G., S.K. and other authors. P.J.H wrote the simulation. A.U., R.R and P.J.H carried out the simulation. A.U and T.G. wrote the manuscript with the input from all authors.

\section{Acknowledgments}
    Use of the Linac Coherent Light Source (LCLS), SLAC National Accelerator Laboratory, is supported by the U.S. Department of Energy (DOE), Office of Science, Office of Basic Energy Sciences (BES) under Contract No. DE-AC02-76SF00515. A.U. acknowledges funding by the Cluster of Excellence 'Advanced Imaging of Matter' of the Deutsche Forschungsgemeinschaft (DFG) - EXC 2056 - project ID 390715994. T.G. acknowledges funding by the European Union (ERC Starting Grant 101040547 - HIGH-Q). Views and opinions expressed are however those of the author(s) only and do not necessarily reflect those of the European Union or the European Research Council Executive Agency. Neither the European Union nor the granting authority can be held responsible for them.
    A. M., D.C., Z.G. and P.F. and acknowledge funding from the U.S. Department of Energy, Office of Basic Energy Sciences, Accelerator and Detector research program. T.D. and J.P.C. were supported by the Chemical Sciences, Geosciences, and Biosciences Division (CSGB), BES, DOE. Z.G. also acknowledges support from the Robert Siemann Fellowship of Stanford University. C.B. and Z.G. acknowledge support from the Swiss National Science Foundation (SNSF) under grant number 200021-197372. D.R acknowledges SNSF grant 200021-232306.

\bibliography{library}
\clearpage

\section {Supplements}
\subsection{Theory}

We employed  Monte-Carlo/Molecular-Dynamics (MC/MD) calculations to simulate the scattering cross sections of the Ne clusters \cite{Ho2016,Ho2020,Kuschel2025} to model the full electron and nuclear dynamics in an atomistic manner during the full duration of the x-ray pulse. In more detail, the interaction of the atom with incident XFEL pulse is treated quantum mechanically with a Monte Carlo method by tracking explicitly the time-dependent quantum transition probability between different electronic configurations. The total transition rate, $\Gamma$, between different electronic configurations $I$ and $J$ is given by
\begin{equation}
\Gamma_{I,J} = \Gamma^{P}_{I,J}+\Gamma^{A}_{I,J}+\Gamma^{F}_{I,J}+\Gamma^{RE}_{I,J}+\Gamma^{EI}_{I,J}+\Gamma^{RC}_{I,J}+\Gamma^{SE}_{I,J}.
\end{equation}
Starting from the ground state of the neutral atom, we include the contribution from photoionization $\Gamma^{P}_{I,J}$, Auger decay $\Gamma^{A}_{I,J}$, fluorescence $\Gamma^{F}_{I,J}$, resonant excitation $\Gamma^{RE}_{I,J}$, electron-impact ionization $\Gamma^{EI}_{I,J}$, electron-ion recombination $\Gamma^{RC}_{I,J}$ and stimulated emission rate $\Gamma^{SE}_{I,J}$. The cross sections and rates are calculated with the Hartree-Fock-Slater model~\cite{Ho2017} with relativistic corrections and spin-orbit coupling in orbital energies.   The stimulated emission rates are derived from Einstein coefficients.  Additionally, a molecular dynamics (MD) algorithm is used to propagate all particle trajectories (atoms/ions/electrons). The cluster dynamics includes electromagnetic forces between the charged particles and van der Waals forces among the neutral atoms.

The scattering response is characterized as a sum of the instantaneous scattering patterns weighted by the pulse fluence, $j_X(\tau,t)$, with FWHM duration $\tau$ and convolved with a Gaussian bandwidth profile, $g(\omega,\omega_x)$, with a central photon energy of $\omega_x$, such that
\begin{eqnarray}
  \label{eq:npSCS}
  \dv{\sigma}{\Omega}(\Vec{q}) = \dv{\sigma_{\mathrm{th}}}{\Omega} \frac{1}{\mathscr{F}} \int_{0}^{+\infty}\!\! \dd\omega \int_{-\infty}^{+\infty}\!\! \dd t \,  g(\omega,\omega_x) j_X(\tau,t) |F_{c}(\Vec{q},t)|^2,
\end{eqnarray}
where $\dd\sigma_{\textrm{th}}/\dd\Omega$ is the Thomson scattering cross section.
\begin{equation}
    \mathscr{F} = \int_{0}^{+\infty}\!\! \dd\omega \int_{-\infty}^{+\infty}\!\! \dd t\,  j_X(\tau, t) g(\omega,\omega_x)
\end{equation}
is the fluence of an XFEL pulse, and $\int_{0}^{+\infty}\!\! \dd\omega\, g(\omega,\omega_x) = 1$. Here $F_{c}(\Vec{q},t)$ is the time-dependent form factor of the target cluster and is modeled as the sum of the form factors of all ions/atoms ($F_{a}(\Vec{q},t)$) and electrons ($F_{e}(\Vec{q},t)$).  Here,   
\begin{equation}
  \label{eq:npFormFactor}
 F_{a}(\Vec{q},t)=\sum_{j=1}^{N_a} f_j(\Vec{q},C_j(t))e^{i \Vec{q} \cdot \Vec{R}_j(t)}\, ,
\end{equation}
where $N_a$ is the total number of atoms/ions, $\Vec{R}_j(t)$, $C_j(t)$ and $f_j(\Vec{q},C_j(t))$ are the position, the electronic configuration and the atomic form factor of the $j$-th atom/ion respectively.   To capture the effect of delocalized electrons in a large cluster, the electrons are assumed to distribute uniformly within the cluster with size $R$, such that  
\begin{equation}
  \label{eq:sphereFormFactor}
 F_{e}(\Vec{q},t)= \frac{3 N_e(t) (\sin(qR)-qR \cos(qR))}{(qR)^3}\, ,
\end{equation}
where $N_e(t)$ is the number of delocalized electrons within the focal region of the x-ray pulse.

\begin{figure}
    \centering
    \includegraphics[width=1\textwidth]{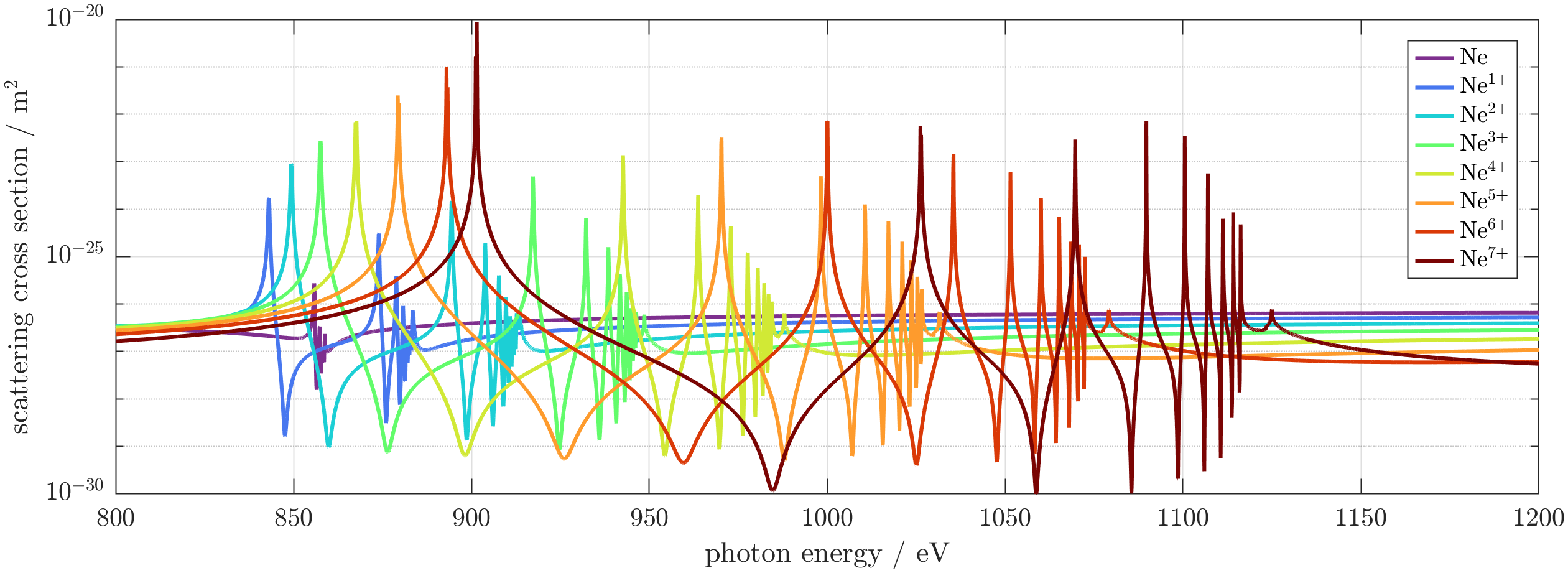}\\
    \includegraphics[width=1\textwidth]{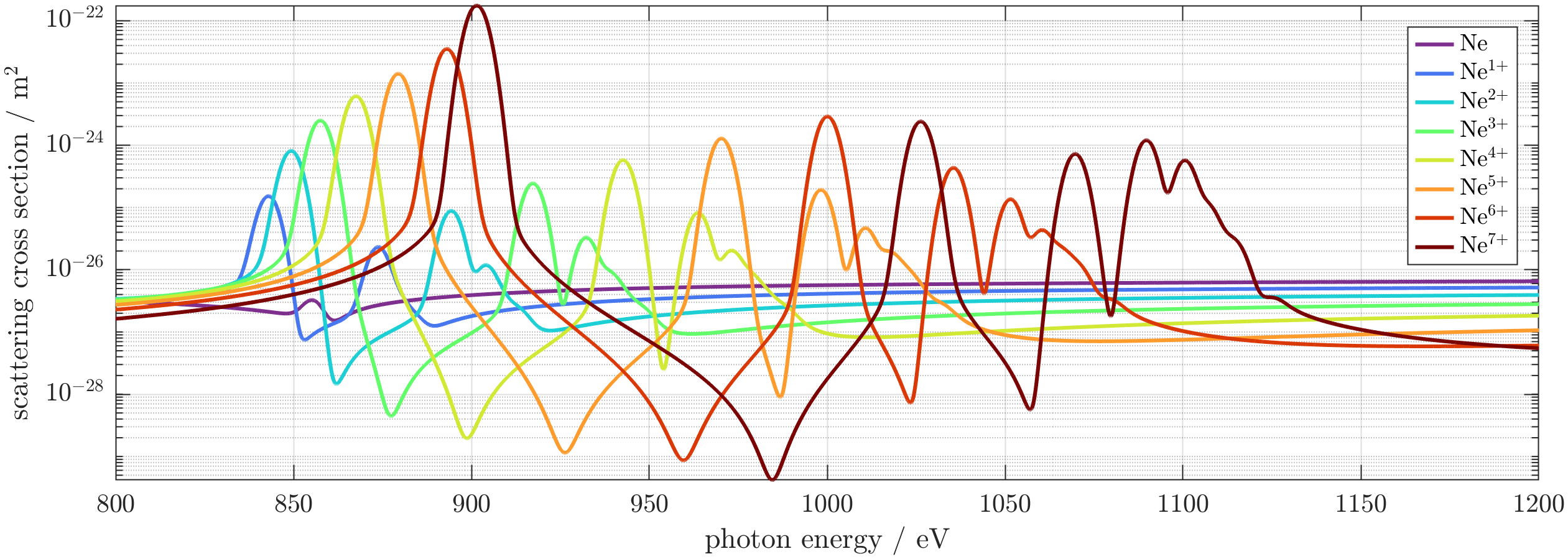}
    \caption{(a) Calculated scattering cross section of different Ne charge states plotted vs incoming x-ray photon energy. (b) Same data as in (a) but convolved with a 6\,eV-Gaussian profile (FWHM) to account for the spectral width of the FEL. 
    }\label{fig:Ne_resonances}
\end{figure}

Our simulation demonstrated in \autoref{fig:theory} overall agrees with the experimental data shown in \autoref{fig:statistics}, especially with the set recorded with sub-fs pulses. Stimulated emission extends the overall lifetime of a small fraction of transient ions (less than 10-15 percent). However, this fraction suffices to increase the image brightness significantly. 

Our simulation indicates that near the Ne K-edge, the average charge state inside the nanoparticle is around Ne$^{4+}$ right after the FEL exposure. During the exposure, there is a charge state distribution ranging from 1+ to up to 5+. In \autoref{fig:Ne_resonances}, the scattering cross sections for different Ne ions are plotted versus incoming photon energy. In case (a), where a bandwidth smaller than 1 eV is assumed, the amplification of diffraction can go up to 6 orders  of magnitude (!). If convoluted with the 6 eV bandwidth of sub-fs pulses, the increase in scattering can still go up by 4 orders of magnitude. Thus, even a small fraction of resonant ions throughout the exposure will increase the total scattering cross section per ion inside the nanoparticle, especially if the lifetime of such resonances is increased through stimulated emission.

There are two discrepancies compared to the experiment, mostly in the plot tracing the 15 fs data set. First, the simulation indicates that there is an elevation of scattering cross section below the Ne edge, which is not present in the data. This behavior might come from a phase shift between the quasi-free electrons and the remaining bound electrons. This shift might be affected by the rapid changes of the overall nanoplasma potential, which is not fully accounted in the simulation. This potential shift through extremely rapid charging might also explain the second deviation: the slight mismatch between experiment and simulation of the photon energies with the highest scattering cross sections. 

In \autoref{fig:Ne_resonances}, the calculated Ne ion cross sections curves up to Ne$^{7+}$ are plotted vs the incoming pulse energy. The neutral case is the black curve. The first maximum of each curve are 1s - 3p transitions, each additional maximum are transitions into higher n shells. We don't expect bound states beyond $n=3$ to survive the potential suppression during nanoplasma formation, thus for our experiment most likely resonances up to 900\,eV play the most crucial role.

\begin{figure}
    \centering
    \includegraphics[width=.6\textwidth]{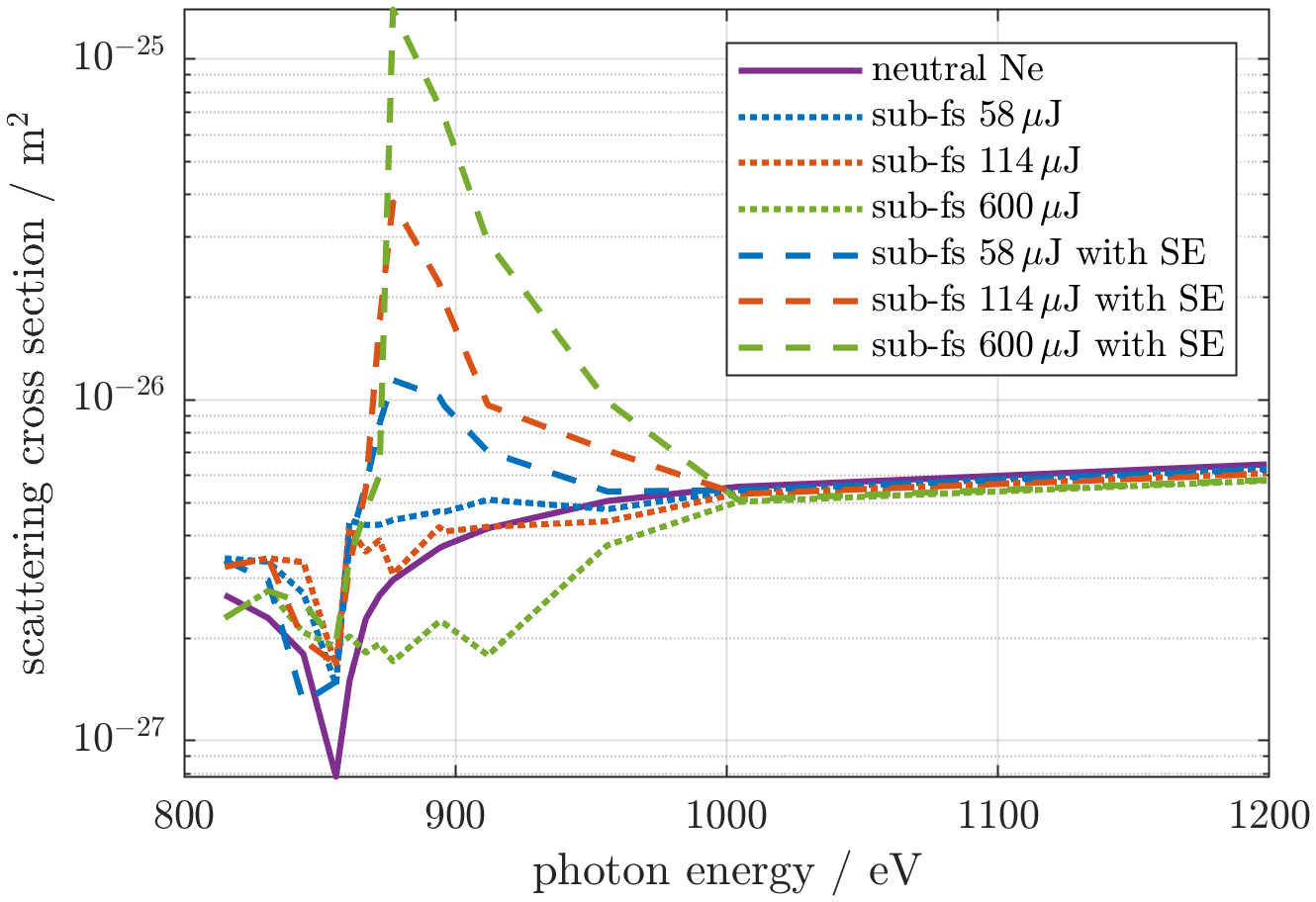}
    \caption{Simulated scattering cross section for neutral Ne atoms inside the nanoparticles (blue solid line), sub-fs pulses with different FEL pulse energies without stimulated emission (dotted lines) and with simulation emission (dashed lines).}\label{fig:Ne_scaling}
\end{figure}
\subsection{Scaling of image brightness with sub-fs pulses}

Our simulation strongly suggests that stimulated emission will play a significant role in improving the brightness and ultimately the resolution of x-ray diffraction images.  In \autoref{fig:Ne_resonances} we plot the results of a thought experiment, where we increase the FEL pulse energy stepwise from \SIrange{0.05}{0.6}{mJ}. Here, the scattering cross sections continue to rise around 900\,eV with the pulse energy if the stimulated emission is included into the simulation. Compared to the neutral curve one can gain almost two orders of magnitude in image brightness beyond scattering from neutral Ne (purple solid curve). Without stimulated emission, we expect to see effects from electronic bleaching even for sub-fs pulses (see dotted green curve). Our thought experiment demonstrates that sub-fs pulses not only beat bleaching by outrunning the nanoplasma formation, but that the diffraction signal from bound electrons becomes dominant when scaling through brightness of images through increase X-ray fluorescence. Bound electrons reflect the pristine structure of the ion positions inside the sample, and thus sub-fs pulses are ideal candidates for true diffraction-before-destruction imaging with atomic resolution.


\subsection{Calculation of the scattering cross-sections from diffraction patterns}

The differential and total scattering cross-sections for an electron (Thomson cross-section) are respectively given by~\cite{Attwood2012}:
\begin{align}
    \frac{\dd \sigma_\mathrm{e}}{\dd \Omega} =r_\mathrm{e}^2 \sin ^2 \Theta \qqq{and}
    \sigma_\mathrm{e} =\frac{8 \pi}{3} r_\mathrm{e}^2
\end{align}
where $\Theta$ is the angle measured from the axis of acceleration $\vb{a}$, accounting for polarization. The differential and total scattering cross-sections for an atom in the case of long wavelength $\left(\lambda \gg a_0\right)$ or small angles $\left(\theta \ll \lambda / a_0\right)$ are respectively given by~\cite{Attwood2012}:
\begin{align}
    \frac{\dd \sigma_\mathrm{a}}{\dd \Omega} = r_\mathrm{e}^2\left|f^0(\omega)\right|^2 \sin ^2 \Theta \qq{and}
    \sigma_\mathrm{a} = \frac{8 \pi}{3} r_\mathrm{e}^2\left|f^0(\omega)\right|^2
\end{align}

The angular distribution of x-rays scattered by a homogeneous sphere with Radius $R$ is given by~\cite{Henke1955, Kirz1995}
\begin{align}
    \frac{\dd \sigma_\mathrm{sph}}{\dd \Omega} &=\left(\frac{\dd \sigma_\mathrm{a}}{\dd \Omega}\right) 8 \pi^3 R^6 n_\mathrm{a}^2 \left(\frac{J_{3 / 2}(qR)}{(qR)^{3/2}}\right)^2 \\
    &= \left(\frac{\dd \sigma_\mathrm{a}}{\dd \Omega}\right) N_\mathrm{a}^2 \left[3 \frac{\sin q R-q R \cos q R}{q^3 R^3}\right]^2
\end{align}
where $q= 2 k \sin (\theta / 2)$, $k = 2\pi/\lambda$ is the angular wavenumber with wavelength $\lambda$, $N_\mathrm{a} = n_\mathrm{a}\,4\pi R^3/3$ is the number of atoms, $J$ is a Bessel function of the first kind and $(\dd \sigma_\mathrm{a} / \dd \Omega)$ is the differential cross-section of a single atom. The total scattered power can be calculated by integrating the radiant intensity over all solid angles $\dd \Omega = \sin{\theta}\,\dd \theta \dd \phi$:
\begin{align}
    P_\mathrm{sca} &= \int I_{r,\Omega}\, \dd \Omega \\
        &= j_\mathrm{sca}^0 \int\limits_0^{2\pi}\int\limits_0^\pi \left[3 \frac{\sin q R-q R \cos q R}{q^3 R^3}\right]^2  \sin{\theta} \,\dd\theta\dd\phi \\
    &= \frac{9\pi}{2k^2R^2} j_\mathrm{sca}^0
\end{align}
On the other hand we have
\begin{align}
     \frac{\dd \sigma_\mathrm{sph}}{\dd \Omega} &= \frac{1}{\abs{S_\mathrm{inc}}} \frac{\dd P_\mathrm{sca}}{\dd \Omega} = \frac{j_\mathrm{sca}(\theta, \phi)}{I_\mathrm{inc}}      = \frac{j_\mathrm{sca}^0}{I_\mathrm{inc}} \left[3 \frac{\sin q R-q R \cos q R}{q^3 R^3}\right]^2 \\ 
    \Rightarrow \sigma_\mathrm{sph} &= \int \frac{j_\mathrm{sca}}{I_\mathrm{inc}} \dd\Omega
    = \frac{j_\mathrm{sca}^0}{I_\mathrm{inc}} \int\limits_0^{2\pi}\int\limits_0^\pi \left[3 \frac{\sin q R-q R \cos q R}{q^3 R^3}\right]^2 \sin{\theta} \,\dd\theta\dd\phi \\
    &= \frac{9\pi}{2k^2R^2}\frac{j_\mathrm{sca}^0}{I_\mathrm{inc}}
\end{align}
Combining the two relations yields
\begin{align}
     \left(\frac{\dd \sigma_\mathrm{a}}{\dd \Omega}\right) &= \frac{j_\mathrm{sca}^0}{I_\mathrm{inc}} \frac{1}{N_\mathrm{a}^2} = r_\mathrm{e}^2 \abs{f^0}^2  \cos ^2 \Theta 
     \xRightarrow{q\to 0} r_\mathrm{e}^2 \abs{f^0}^2 = \frac{j_\mathrm{sca}^0}{I_\mathrm{inc}} \frac{1}{N_\mathrm{a}^2} \\
\end{align}
allowing to compare the scattering cross-section per atom with tabulated atomic values $\sigma_\mathrm{a}$ if irradiance $I_\mathrm{inc}$ and scattered radiant intensity $j_\mathrm{sca}^0$ are known. 

\subsection{Experimental Details}

The experiment was performed at the time-resolved atomic, molecular and optical science (TMO) instrument~\cite{Osipov2018lamp, Walter2022TMO} of the linac coherent light source (LCLS)~\cite{Emma2010}. X-ray pulses were produced in two operation modes, by (a) the 'conventional' self-ampliﬁed spontaneous emission (SASE) scheme with electron bunch pulse lengths of 15\,fs to 20\,fs, and (b) using the novel x-ray laser enhanced attosecond pulse generation (XLEAP) technique~\cite{Duris2020}, delivering near-Fourier limited sub-femtosecond pulses~\cite{Duris2020}. The actual x-ray pulse length is around two/thirds of the electron bunch, on average smaller than 15 fs. \autoref{tab:pulseTypes} shows an overview of the x-ray pulse parameters. 

\begin{table}
    \centering
    \caption{The three different pulse types used throughout the present experiment. The pulse duration (FWHM), pulse energy (mean $\pm$ standard deviation) and energy bandwidth (mean $\pm$ standard deviation of FWHM at different energies) are denoted as $\tau_\mathrm{FWHM}$, $\langle E_\mathrm{p} \rangle \pm \sigma_\mathrm{p}$ and $\langle \Delta E \rangle \pm \sigma_{\Delta E}$, respectively.
    }\label{tab:pulseTypes}
    \begin{tabularx}{.48\textwidth}{@{\extracolsep{\fill}}lccc}
        \toprule
        pulse type & $\tau_\mathrm{FWHM}$ & $\langle E_\mathrm{p} \rangle \pm \sigma_\mathrm{p}$ & $\langle \Delta E \rangle \pm \sigma_{\Delta E}$ \\
        \midrule
        XLEAP & \SI{\sim 0.3}{fs} & $(62 \pm 26)\,\si{\micro\joule}$ & $(6.4 \pm 0.7)\,\si{eV}$ \\
        SASE atten & 10--\SI{15}{fs} & $(114 \pm 20)\,\si{\micro\joule}$ & $(5.6 \pm 0.9)\,\si{eV}$ \\
        SASE full & 10--\SI{15}{fs} & $(600 \pm 90)\,\si{\micro\joule}$ & $(5.6\pm 0.9)\,\si{eV}$ \\
        \bottomrule
    \end{tabularx}
\end{table}

A pair of Kirkpatrick--Baez mirrors focused the x-ray beam to \SI{\sim 1.2}{\micro\metre} in diameter (full with at half maximum, FWHM) in the interaction point (IP). \autoref{fig:exp} shows a sketch of the experimental layout. 

Clusters were produced by expansion of pre-cooled Ne gas into a vacuum chamber. The source employed a cryogenic Even--Lavie valve~\cite{Even2016} with a narrow conical nozzle ($\varnothing{}\SI{100}{\micro\meter}$, \ang{2} half-opening angle) and operated at \SI{60}{Hz} and \SI{60}{\micro\second} opening time. The stagnation pressure was set to \SI{2}{MPa} at stagnation temperatures of \SIrange{43}{47}{K}. For details on the cluster source setup please see Refs.~\cite{Langbehn2021, Langbehn2018}. The neon cluster beam passed through a \SI{1}{mm} skimmer and a second piezo-driven variable-gap skimmer, allowing for hit rate reduction and single-shot single-particle operation.

X-rays diffracted by individual neon clusters were recorded with an ePix100 detector~\cite{Dragone2014}, containing $768 \times 704$ pixels ($50\times\SI{50}{\micro\metre^2}$ pixel size) located \SI{0.395}{m} after the IP. The x-ray photon was mounted shifted to one side of the beam stop, in order to capture lowest order diffraction. The  detector covered diffraction angles up to \ang{4.9} (edge), the Si beam stop blocked signal from angles below \ang{0.7} and the direct FEL beam. An ion time-of-flight (iTOF) spectrometer with a 1\,mm entrance slit was installed above the IP, recording coincident ion spectra. The slit prevents ions from outside of the Rayleigh length. See Ref.~\cite{Ferguson2016transient} for details on the iTOF geometry and tranmission function.

The ePix100 detector was movable and was driven completely out of the beam in order to complete the absolute photon energy calibration. A Fresnel zone plate (FZP) spectrometer (range: \SI{821}{eV} to \SI{886}{eV}), located behind the diffraction detector, was used for FEL photon energy calibration, see Ref.~\cite{Larsen2023} for details. FEL spectra were recorded for photon energies between 831\,eV and 877\,eV for XLEAP pulses, and between 844\,eV and 872\,eV for SASE pulses, respectively (calibrated values). A linear fit was used to calculate calibrated photon energies shot-to-shot from the nominal machine parameters. The FZP spectrometer was calibrated on the Ni L2 (870.0\,eV) and L3 (852.7\,eV) edges by tuning the FEL to the respective energy, and recording spectra without and with a thin Ni absorption foil, for SASE and XLEAP operation, respectively.

We recorded a reference data set with unattenuated 15 fs FEL pulses, which acted as a check for consistency and marked in the table as SASE high. This data is not shown in the paper for brevity reasons as no comparable sub-fs pulses were available. 


\

\subsection{Data Processing}

\begin{figure}[ht]
    \centering
    \includegraphics[width=1\textwidth]{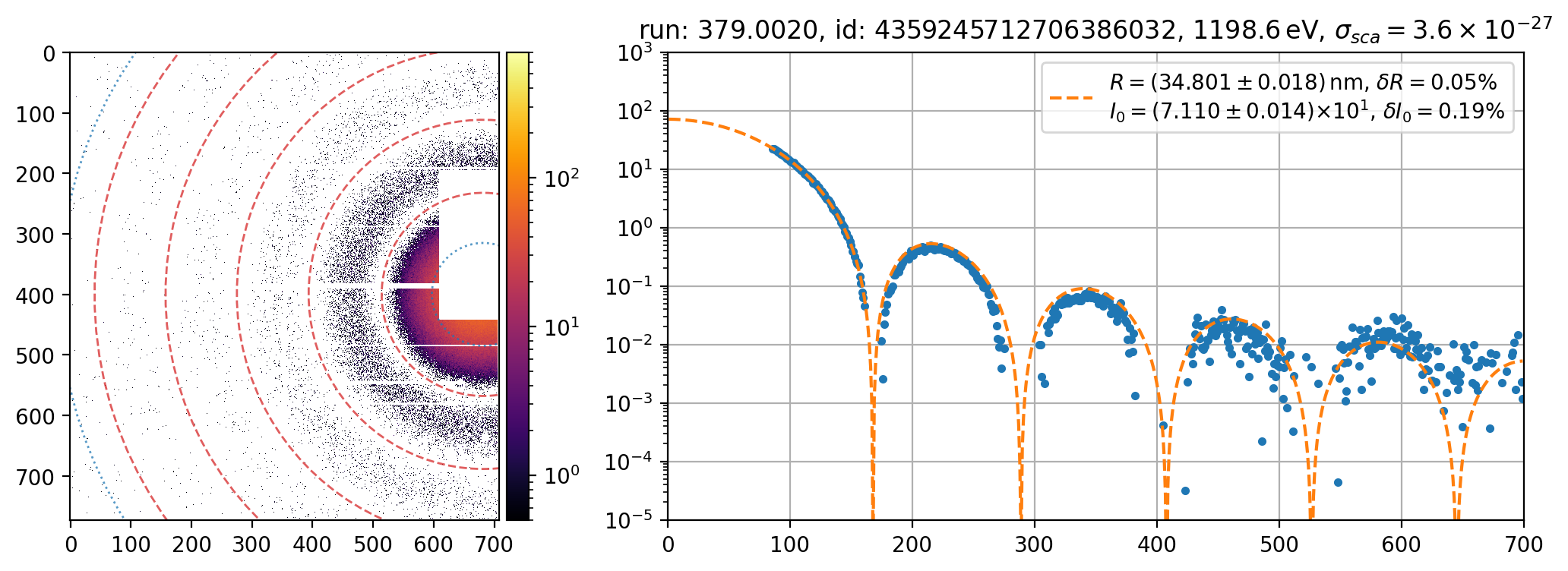}
    \caption{Example diffraction image from an individual nanoparticle (left, x,y axis are pixel numbers) and the corresponding one dimensional plot of the diffraction (right, x-axis is pixel number, y-axis signal strength). The diffraction signal (blue dots) is fitted by the shape of a phase-shifting sphere as described below. The fit is shown in as dashed line, the locations of fitted minima is indicated through dashed line in the 2D diffraction image.}\label{fig:example}
\end{figure}

\subsubsection{Diffraction Images}

In a first step, the diffraction detector images were masked for inactive, dead and hot pixels. In addition the pedestal, background and common mode corrections were applied. Each experimental run contained approximately 18000 FEL shots, from which the brightest 500 were selected for further analysis. 

\begin{figure}
    \centering
    \includegraphics[width=1\textwidth]{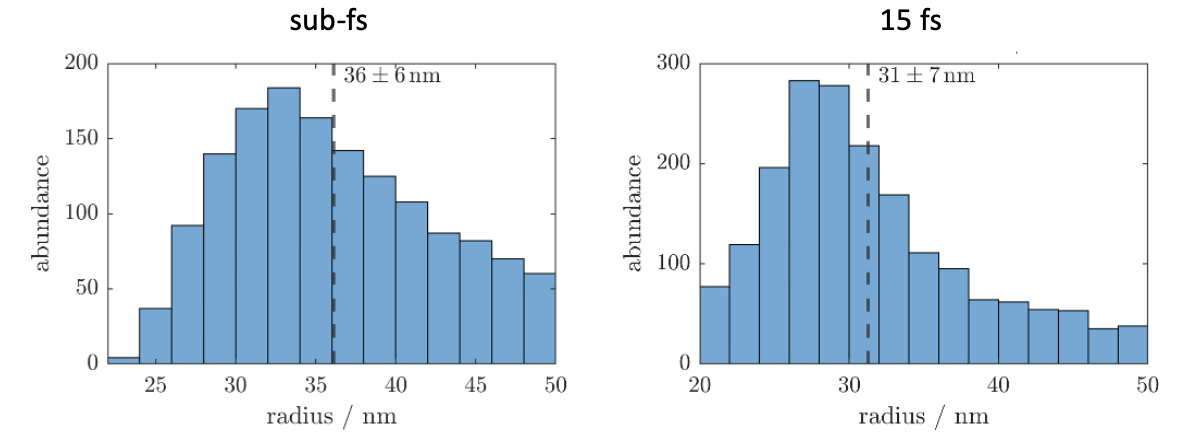}
    \caption{Radius histograms of the clusters (filtered for 20\,nm to 50\,nm) used throughout the present analysis. The different distributions are presumably resulting from differences in the detection efficiency due to different photon fluences of the individual pulse types. The (mean $\pm$ standard) deviation values for the three pulse types are \SI{36 \pm 6}{nm} for XLEAP and \SI{31 \pm 7}{nm} for SASE attenuated. 
    }\label{fig:Sizes}
\end{figure}

The nanoparticle size was extracted from the diffraction image as shown in \autoref{fig:example} and described below.
The radial profiles were fitted to the diffraction profile of a non-absorbing sphere with constant density~\cite{Guinier1955}
\begin{equation}
    j_\mathrm{sph} (q) = j_\mathrm{sph}^0 \left[3 \frac{\sin q R-q R \cos q R}{q^3 R^3}\right]^2 ,
\end{equation}
yielding the cluster radius $R$ and scattered radiant intensity into forward direction
\begin{equation}
    j_\mathrm{sph}^0 = I_\mathrm{inc} r_\mathrm{e}^2 \left| f^0 (\omega) \right|^2 N_\mathrm{a}^2
\end{equation}
as fit parameters. The magnitude of the elastic momentum transfer vector is given by $q = 2 k\,\sin(\theta/2)$, with the angular wave number $k = 2\pi/\lambda$ and the diffraction angle $\theta$. Here, $I_\mathrm{inc}$ describes the irradiance/incident intensity (power per unit area), $r_\mathrm{e}$ the classical electron radius, $f^0 (\omega)$ the atomic scattering factor, and $N_\mathrm{a} = 4\pi R^3 n_\mathrm{a}/3$ the number of atoms in the cluster, where $n_\mathrm{a} = \SI{43}{nm^{-3}}$ is the number density of neon at the triple point~\cite{Klein1997}. The fit parameter $j_\mathrm{sca}^0$ was additionally corrected for the wavelength-dependent quantum efficiency $D_\mathrm{QE}$ of the scattering detector, which increases approximately linearly from $11\%$ to $44\%$ for photon energies ranging from \SI{800}{eV} to \SI{1200}{eV}. All processed images were checked for the alignment of the fit to the data. Multiple particle hits or fits greatly diverging from a sphere were sorted out. Additionally, only clusters with radii between \SI{20}{nm} and \SI{50}{nm} were selected for further analysis, to improve comparability and ensure that absorption effects are minute. Size distributions of the evaluated data set can be found in \autoref{fig:Sizes}. The scattering cross section can be calculated as follows:
\begin{equation}
    \sigma_\mathrm{sca} = \frac{8\pi}{3} r_\mathrm{e}^2 \left| f^0 (\omega) \right|^2 = \frac{8\pi}{3} \frac{j_\mathrm{sph}^0}{I_\mathrm{inc} N_\mathrm{a}^2} .
\end{equation}

\begin{figure}
    \centering
    \includegraphics[width=\plotwidth]{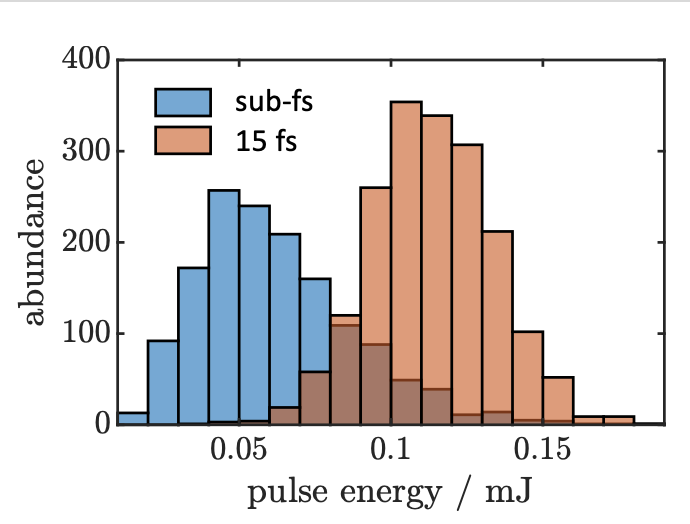}
    \caption{Pulse energy histograms (on target) for the analyzed data set. Only shots with good hits and near-spherical nanoparticles are shown. The distribution of sub-fs pulses has a tail towards higher pulse energies, which is not present in the Gaussian-like distribution of 15 fs pulses. In order to cover the most intense sub-fs pulses, the average pulse energy of 15 fs pulses was shifted to higher pulse energies  }\label{fig:PulseEnergies}
\end{figure}

The pulse energies were recorded using a gas monitor detector (GMD) and corrected for a beamline transmission of 60\%. Histograms of the pulse energies of the evaluated data set are shown in \autoref{fig:PulseEnergies}. The incident intensity was calculated with $I_\mathrm{inc} = \frac{2 P_\mathrm{tot}}{\pi w_0^2}$ from the total pulse power $P_\mathrm{tot}$, assuming a Gaussian beam profile with a beam waist of $w_0=\frac{d_\mathrm{FWHM}}{\sqrt{2 \ln 2}}$ and for a beam diameter of $d_\mathrm{FWHM}$. 

Since only a minor fraction of the clusters were hit by the central beam, while most clusters were illuminated by the beam wings, the calculated scattering cross-section is underestimated in most cases. Therefore, only the top 5\% of the calculated values for $\sigma_\mathrm{a}$ are used to estimate the average atomic scattering cross-section $\langle \sigma_{a}^j \rangle_{5\%}$ for each pulse type $j \in \left\{ \text{XLEAP, SASE atten} \right\}$.
\begin{align}
    \langle \sigma_{a}^j \rangle_{5\%} = \mathrm{mean} \left[\{\sigma_{a, i}^j\} \geq \eta_{95\%} (\{\sigma_{a, i}^j\} ) \right]
\end{align}
with $\eta_{95\%}(\{\sigma_{a, i}^j\})$ denoting the 95th percentile of the set of shots $\{\sigma_{a, i}^j\}$. The beam diameter is known to be $\gtrsim$ 1.2 $\mu m$ from wavefront measurements~\cite{Walter2022}. We confirmed the value based on linear regime scattering below the Ne absorption edge. For the clusters which were illuminated with sub-fs pulses below the edge (around \SI{815}{eV})  approximately one photon per atom may be absorbed and no transient resonances are present. Thus, the diffraction should follow literature values. Fitting $\langle \sigma_{a}^\mathrm{X} \rangle_{5\%} (\SI{815}{eV})$ to the literature value of the linear atomic scattering cross-section~\cite{Henke1955} yields a beam diameter of $d_\mathrm{FWHM} = \SI{1.17}{\micro m}$, which is in good agreement with the expected value from the wavefront measurements.


\subsubsection{Time-of-Flight Spectra}

 The ion time-of-flight (iTOF) spectra were recorded coincidently with the scattering patterns using a spectrometer described in detail in Ref.~\cite{Ferguson2016Diss}. Due to an impedance mismatch at the digitizer, the signal was distorted by reflections. In a first step, the signal was corrected by subtracting a moving background (using the \textsc{Matlab}~\cite{MATLAB} \mbox{'msbackadj'} function with parameters: window size = 50 samples, step size = 25 samples, quantile = 0.05, regression method = 'pchip'). The flight times were calibrated using peaks of $^{20}$Ne$^{Q+}$ and $^{22}$Ne$^{Q+}$ with charges $Q$ ranging from 2 to 8 from spectra of dilute neon gas. While iTOF spectra of neon gas show sharp peaks for the element specific $m/Q$ values, the spectra of clusters are broadened by the kinetic energies the ions gain from interaction with the electric field of the other cluster fragments. In neon gas only ionization states up to Ne$^{8+}$ is accessible by photon energies below \SI{1200}{eV} \cite{Young2010}. Therefore ions with shorter flight times indicate either additional ionization channels within the cluster, e.g. through electron impact ionization in the plasma; or (probably mainly) kinetic energy from the acceleration resulting from larger space charges. Both of these indicate increased absorption in the cluster, which is why the average cluster charge calculated from the average flight time is used as an indicator for absorption, even if it does not necessarily directly map to real ionization states. In addition, one should note, that ion spectra are signatures of nanoplasma, which had microseconds of recombination after the FEL has passed. In previous studies, ion spectra was reliably used to estimage the overall absorbed energy from the FEL pulses, but not for the measurement of charge states during the FEL exposure \cite{Gorkhover2012, Gorkhover2016, Rupp2020}.


\end{document}